\documentclass[a4paper]{jpconf}
\usepackage{graphicx}
\usepackage{hyperref}

\newcommand{\gerda}{\textsc{Gerda}}

\newcommand{\gelatio}{\textsc{Gelatio}}

\begin{document}
\title{Off-line data processing and analysis for the GERDA experiment}

\author{M~Agostini$^{1}$, L~Pandola$^2$ and P~Zavarise$^{2,3}$}
\address{$^1$~Physik Department E15, Technische Universit\"at M\"unchen,
Garching, Germany}
\address{$^2$~INFN, Laboratori Nazionali del Gran Sasso,
Assergi (AQ), Italy}
\address{$^3$~Dipartimento di Fisica, Universit\`a dell'Aquila,
L'Aquila, Italy}

\ead{matteo.agostini@ph.tum.de, pandola@lngs.infn.it, zavarise@lngs.infn.it}

\begin{abstract}
\gerda\ is an experiment designed to look for the neutrinoless double beta
decay of $^{76}$Ge. 
The experiment uses an array of high-purity germanium detectors, enriched in
$^{76}$Ge, directly immersed in liquid argon.
\gerda\ is presently operating eight enriched coaxial detectors (approximately 15 kg of
$^{76}$Ge) and about 30 new custom-made enriched BEGe detectors will be deployed
in the next phase (additional 20\,kg of $^{76}$Ge).
The paper describes the \gerda\ off-line analysis of the high-purity germanium
detector data.
Firstly we present the signal processing flow, focusing on the digital filters
and on the algorithms used.
Secondly we discuss the rejection of non-physical events and the
data quality monitoring.
The analysis is performed completely with the \gerda\ software
framework (\gelatio), designed to support a multi-channel processing and to 
perform a modular analysis of digital signals.
\end{abstract}

\section{Introduction}
\gerda\ \cite{gerda} is a low-background experiment searching for the
neutrinoless double beta decay of $^{76}$Ge, using an array of bare high-purity
germanium (HPGe) detectors isotopically enriched in $^{76}$Ge. 
The detector array is operated directly in ultra radio-pure liquid argon, allowing a
substantial background reduction at the $Q_{\beta\beta}$-value of $^{76}$Ge with
respect to the previous experiments~\cite{hm}.
In the present phase (Phase\,I) eight enriched coaxial detectors are being used, 
totaling approximately 15 kg of $^{76}$Ge.
In Phase\,II, about 30 new custom-made enriched BEGe
detectors~\cite{bege} will be deployed (additional $\sim$20\,kg of $^{76}$Ge).
The experiment is located in the underground Laboratori Nazionali del Gran Sasso
of the INFN (Italy).

The background suppression in \gerda\ is achieved by the specific innovative 
design (namely, detectors operated naked in a cryogenic liquid) and by a strict  
material selection. In addition, part of the remaining background
events can be identified by an off-line analysis of the HPGe detector signals,
i.e. detector anti-coincidence and pulse shape discrimination
techniques~\cite{psd}.
%

For this purpose, a software framework (\gelatio) for advanced digital signal
processing and analysis  has been recently developed~\cite{gelatio}. 
It is implemented in C++ and is based on the \textsc{MGDO} library~\cite{mgdo}.
The framework is designed to support a multi-channel data processing and a 
modular analysis of digital signals.  
Signals are analyzed by using chains of modules completely customizable by
the user.
Each module handles a precise and self-consistent task of the signal
processing and is implemented as a dedicate C++ class.
The output of the modules, which is either a scalar parameter (e.g. the
amplitude of the signals) or a shaped trace, can be used as input for other
modules and/or stored to disk.

The framework was used for the reference analysis of the data acquired in 
the \gerda\ commissioning phase (from June 2010 to October 2011), when up to
seven HPGe detectors have been operated simultaneously.
The commissioning data were used as a benchmark to validate \gelatio\ against
other independent analysis codes and to prove its suitability for
the use in the \gerda\ Phase\,I.

This paper describes the basic off-line analysis of the \gerda\ data performed
with the \gelatio\ framework.
In section~\ref{dsp} we present the flow of the signal processing and analysis
along the module chains and the shaping algorithms.
Then, in section~\ref{filter}, we discuss the identification of non-physical
events or of signals not properly processed along the analysis pipeline. Also, the
monitoring of the data quality will be described.
Finally, summary and conclusions are presented in section~\ref{last}.

\section{Signal processing flow}\label{dsp}
The charge pulses from the HPGe detectors
operated in \gerda\ are digitized by
14-bit flash-ADCs~\cite{daq} (FADC) running at 100\,MHz sampling rate. 
For each event, the FADC computes in run-time two traces that are eventually
written to disk. 
The first trace is sampled at 100\,MHz and is 4\,$\mu$s long 
(high-frequency-short trace). 
It includes the signal leading edge and it is used for identifying
background events through pulse shape discrimination techniques. 
The second trace has a sampling frequency of 25\,MHz and is
160\,$\mu$s long (low-frequency-long trace). 
It is used for those operations, as energy reconstruction, which involve 
the integration of the signal.

\begin{figure}[tb]
   \begin{center}
   \includegraphics[width=0.9\textwidth]{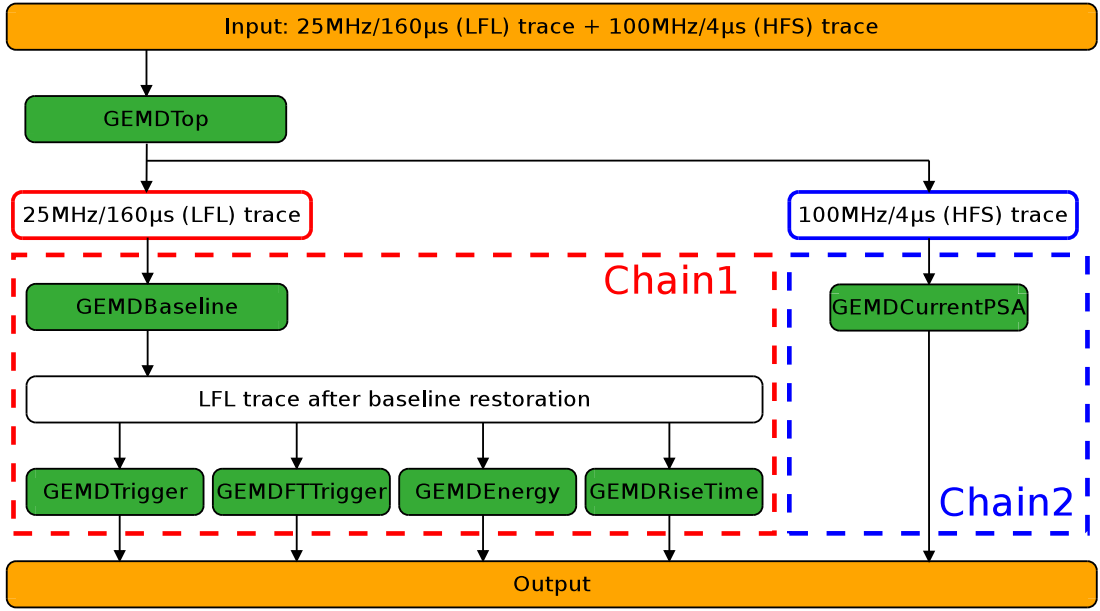}
   \end{center}
   \caption{\small Flow chart of the signal processing. The two traces saved by
   the digitizer are processed along two different chains of \gelatio\ analysis modules. 
   The
   low-frequency-long (LFL) trace is used for reconstructing the energy, the
   trigger and the rise time. The high-frequency-short (HFS) trace is used for
   pulse shape analysis.}
   \label{calflow}
\end{figure}

The two traces are processed along different chains of \gelatio\ modules, as shown in
\figurename~\ref{calflow}.
The first module is GEMDTop. It takes care of extracting the traces from the
input file and making them available to the other modules. It also checks, and
possibly changes, the pulse polarity in order to always have positive-polarity
pulses. The output traces are the starting point for two chains: the
low-frequency-long trace is processed along Chain1 while 
the high-frequency-short trace along Chain2.

Chain1 starts with GEMDBaseline. 
This module analyzes the baseline of the signal by computing the average
value, the rms and the linear slope in the pre-trigger region. 
In addition, the module performs a baseline restoration --- a subtraction of the 
average baseline value to the trace --- and provides the new signal  
to the other modules:
\begin{itemize}
   \item GEMDTrigger. The module implements a leading-edge discriminator with
      threshold defined dynamically as three times the rms of the signal
      baseline. 
      After the trigger, the signal has to remain above threshold for 
      at least 40\,$\mu$s, 
      otherwise the trigger is rejected.
      Before searching for the trigger, the pulse is integrated using a
      160\,ns moving average filter in order to reduce the high-frequency noise.%
   \item GEMDFTTrigger. 
      The module applies to the input signal a 1.5\,$\mu$s moving
      differentiation filter and a 1\,$\mu$s moving average filter for noise
      reduction (see
      \figurename~\ref{shaping1}).
      The resulting trace has a peak for each sharp variation of the signal
      (such as the leading edge of a pulse) 
      and is analyzed by a leading-edge discriminator. 
      The peak width is similar to the size of the moving differentiation and
      was chosen to maximize the pile-up identification efficiency and to
      avoid the mis-identification of highly-multiple-site events.
      The number and the position of the peaks are estimated by 
      applying a leading-edge discriminator, whose threshold is 
      four times the rms of the baseline. 
      After this condition is met, the signal must remain above the threshold for
      at least 1\,$\mu$s.
      While GEMDTrigger is tuned to determine the trigger position with high
      precision and stability (it requires that the signal remains above the
      threshold for 40\,$\mu$s), 
      this module is important to identify 
      events with multiple physical signals occurring within the same trace.
   \item GEMDEnergyGauss. The module reconstructs the event energy using an
      approximate Gaussian filter~\cite{dspguide}. 
      The pulse is differentiated by a moving differentiation filter
      and then integrated 15 times by a moving average filter to achieve an
      approximated Gaussian shape\footnote{%
      Historically, the energy reconstruction filters for $\gamma$-ray spectroscopy
      also perform a deconvolution of the exponential function
      which is folded in the signal by the charge sensitive pre-amplifier.
      However, the approach suggested in this paper was found to provide better
      results on the \gerda\ data with respect to the usual filters. 
      }. 
      The energy information is eventually stored
      in the maximum amplitude of the quasi-Gaussian pulse.
      The width of the moving filters has been set to 10\,$\mu$s in order to minimize 
      losses due to ballistic effects.
      The intermediate steps of the shaping are shown in
      \figurename~\ref{shaping2}.
      \begin{figure}[tbp]
      \begin{minipage}[t]{0.48\textwidth} 
      \begin{center}
      \includegraphics[angle=270,width=\textwidth]{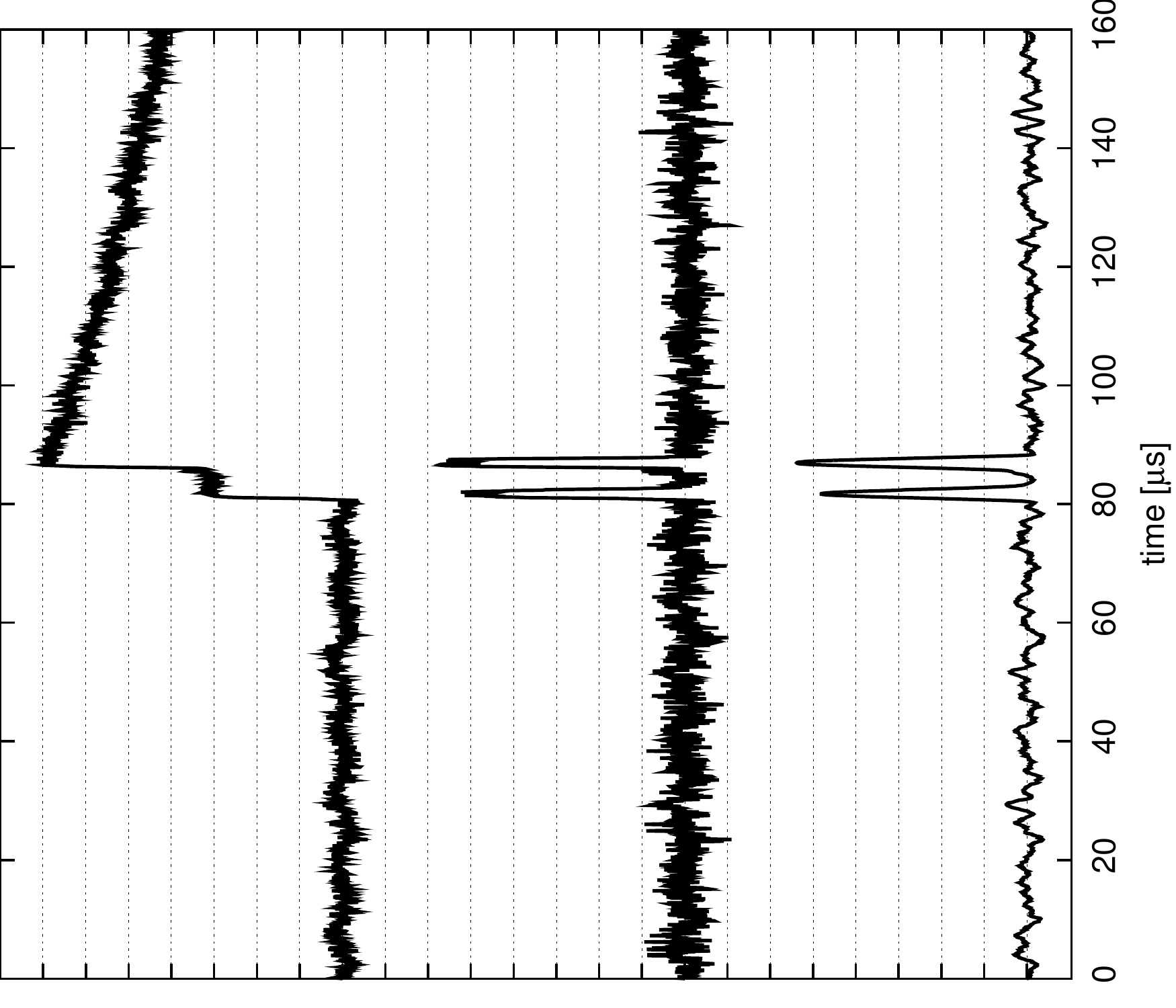}
      \caption{\small
      Digital signal processing performed by GEMDFTTrigger.
      The incoming signal~(top trace) is differentiated (middle) and then
      integrated.
      Each sharp variation of the incoming pulse creates a peak in the output trace
      (bottom trace).}
      \label{shaping1}
      \end{center}
      \end{minipage}
      \hspace{0.02\textwidth}
      \begin{minipage}[t]{0.48\textwidth} 
      \begin{center}
      \includegraphics[angle=270,width=\textwidth]{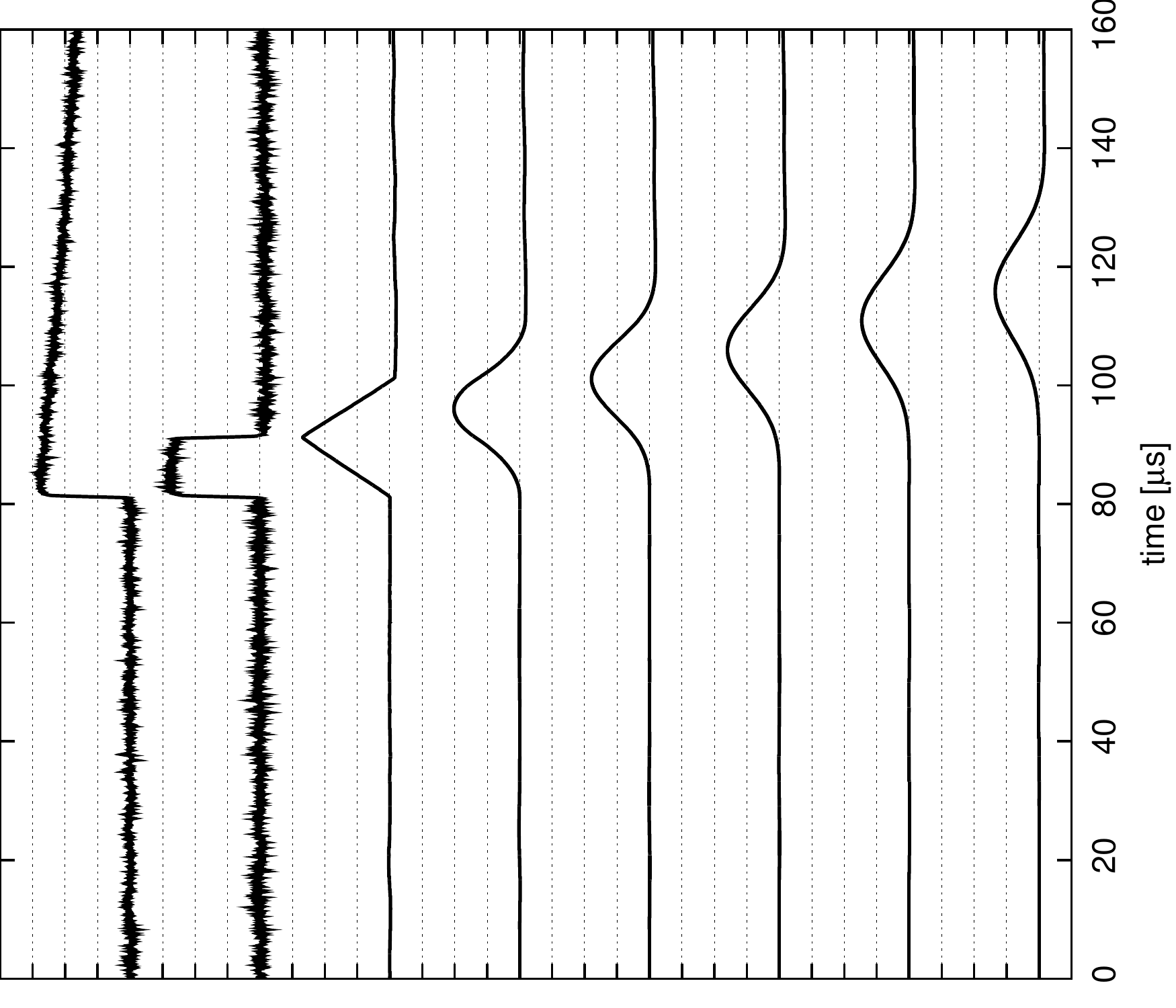}
      \caption{\small 
      Digital signal processing performed by GEMDEnergyGauss.
      The incoming signal (top trace) is differentiated
      (second trace) and then integrated several times (following traces)
      by a moving average filter.}
      \label{shaping2}
      \end{center}
      \end{minipage}
      \end{figure}
   \item GEMDRiseTime. The module computes the rise time between 10\% and 90\% of 
     the maximum amplitude of the pulse.
      The maximum amplitude is computed as the difference between the maximum of the
      pulse and the average baseline value. 
      Then, the first samples below the 10\% and 90\% of the maximum amplitude
      are found by moving backwards from the position of the maximum.
\end{itemize}
The second chain is used to evaluate parameters relevant for pulse shape
discrimination techniques and it will be better defined during the future data
taking.
The chain presently includes only one module, GEMDCurrentPSA, which computes the
current pulse as the derivative of the signal and then extracts the basic
features of the current peak, like rise time, width and area.

\section{Data selection and monitoring}\label{filter}
In the \gerda\ data sets there are two main classes of signals that have to be
identified and tagged: 1) signals corrupted or produced by non-physical events, i.e. 
discharges, cross-talk, pick-up noise; 2) signals which are not properly
processed along the analysis pipeline, as pile-ups and accidental coincidences.

\begin{figure}[t]
\begin{minipage}[t]{0.48\textwidth} 
\begin{center}
\includegraphics[angle=270,width=\textwidth]{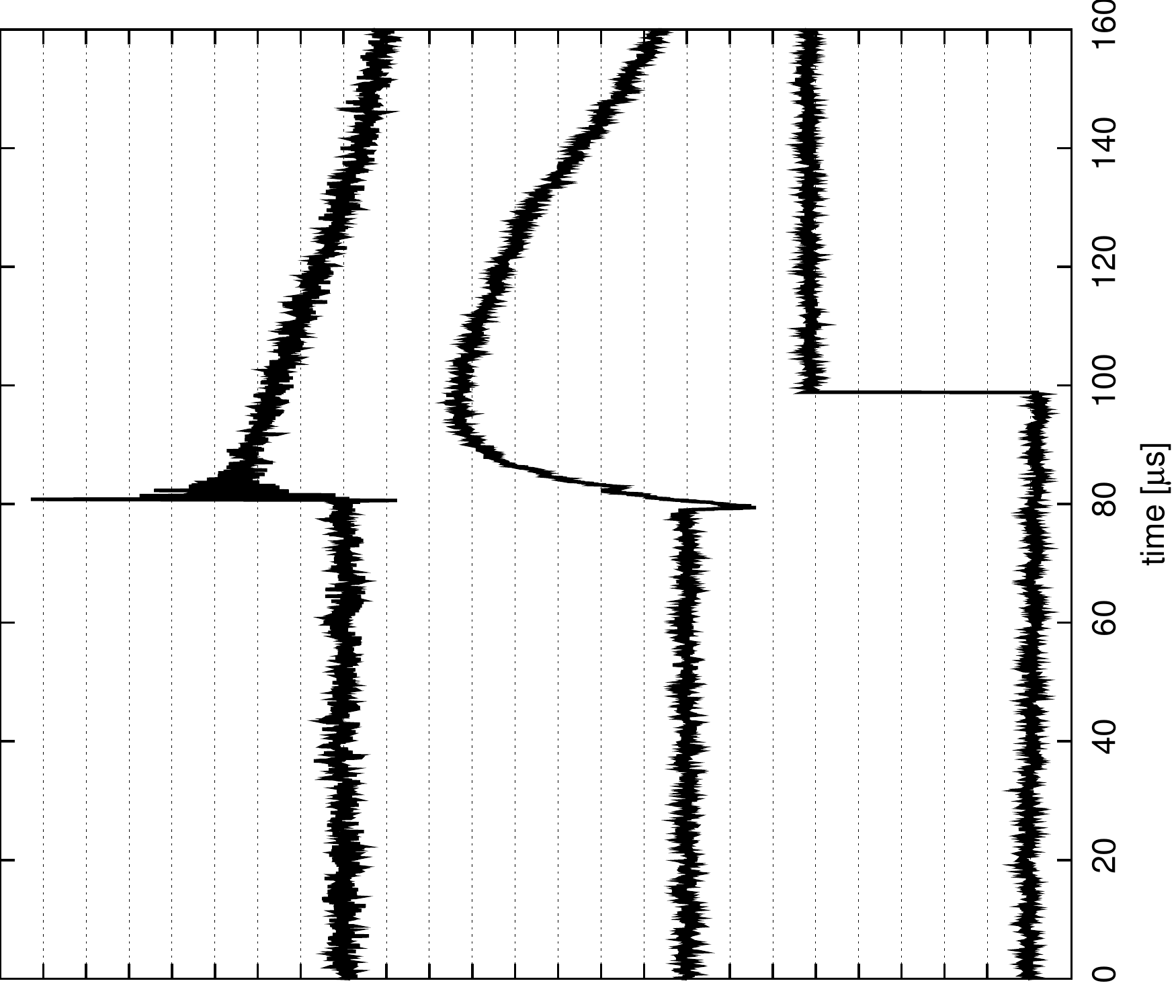}
\caption{\small
Illustrative traces generated by non-physical events. Note that these 
signals do not have the typical exponential decay tail after the leading edge.
}
\label{filter1}
\end{center}
\end{minipage}\hspace{0.02\textwidth}
\begin{minipage}[t]{0.48\textwidth} 
\begin{center}
\includegraphics[angle=270,width=\textwidth]{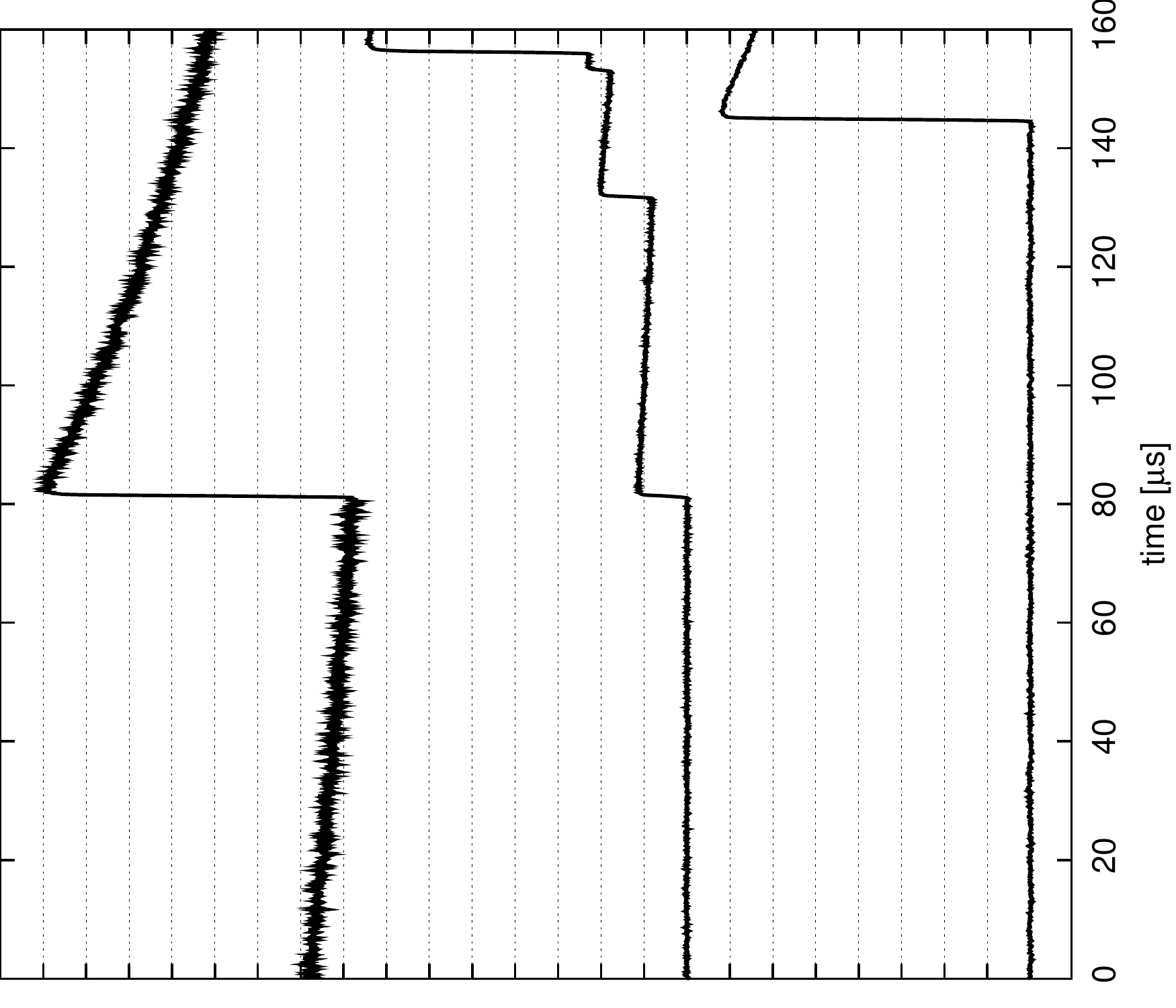}
\caption{\small
Example of traces generated by pile-up signals (top-middle trace) and
accidental coincidences (bottom).
}
\label{filter2}
\end{center}
\end{minipage}
\end{figure}

The first class includes signals with
anomalous shape, wrong polarity, extremely short/long rise time or exceeding
the dynamic range of the FADC (see \figurename~\ref{filter1}).
To identify these events a sequence of cuts based on four parameters is applied.
The first parameters are the trigger position computed by GEMDTrigger and the
time position of the maximum amplitude of the Gaussian pulses (maxAmpTime)
computed by GEMDEnergyGauss. 
If the signal has a leading edge at the proper position,
followed by an exponential decay tail due to the charge sensitive pre-amplifier,
then the trigger has to be reconstructed roughly in the center of the trace and
maxAmpTime has to be in a well-defined range.
The third parameter is the 10-90\% rise time  which can be used to identify
signals that are extremely fast or slow, and hence inconsistent with well-behaved 
physical events. Finally, signals that saturate the dynamic 
range of the FADC are identified by scanning the individual 
samples of the traces.

The second class includes signals generated by the superimposition of multiple physical
pulses, or having the leading edge not aligned with the center of the
trace (see \figurename~\ref{filter2}).
These signals can be identified using the baseline slope computed by
GEMDBaseline, the number of pulses provided by GEMDFTTrigger and the trigger
position yielded by GEMDTrigger.
Their amount is proportional to the event rate and can reach up to 15\% in
the calibration data sets, while it is usually negligible in the physics data sets. 
Therefore, the identification of these signals is a critical issue to
derive from the calibration data a sample of events which is as similar 
as possible to the physics run data.

\begin{figure}[t]
\begin{center}
   \includegraphics[angle=270,width=\textwidth]{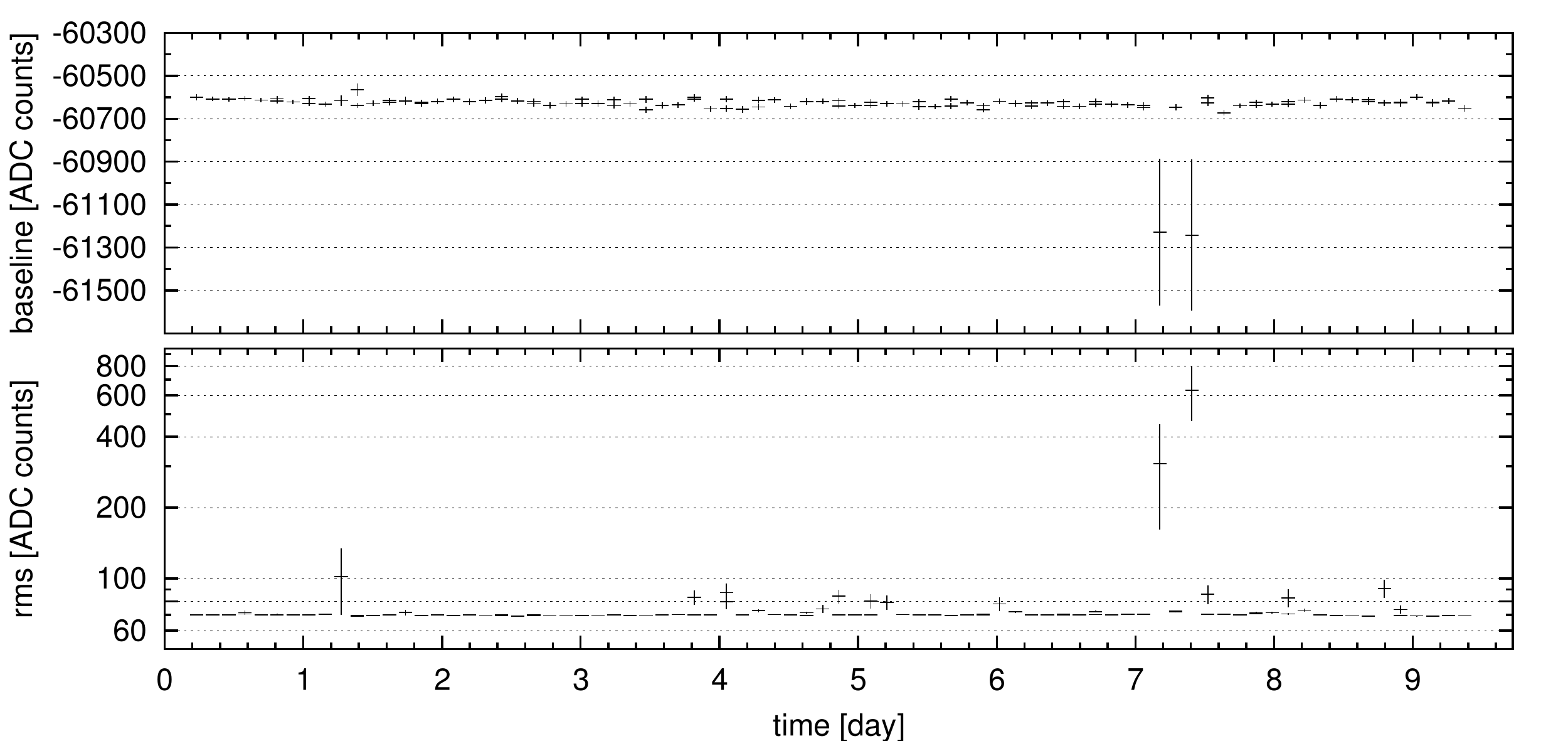}
\end{center}
\caption{\small
Average value (top panel) and rms (bottom panel) of the signal baseline vs. time,  
for a detector operated in \gerda\ during a 10-day run. 
The bin content represents the mean value of the parameter and the error
bars are related to the width of the distribution. The bins are 2~hours wide. 
}
\label{profile}
\end{figure}

Besides cuts for removing undesirable classes of signals, 
there are also parameters which can be used to monitor the quality of the data taking
and the stability of the set-up, the most important being the average value
and the rms of the baseline. These parameters are sensitive to noise changes and to 
gain variations in the read-out chain. 
\figurename~\ref{profile} shows these parameters as a function of time for a 10-day 
commissioning run. 
The parameters are stable over the whole data taking, except for a few hours
during day 7. These instabilities can be correlated with hardware operations in
the set-up and the corresponding data can be removed by applying a cut on the
two parameters.

\begin{figure}[tbp]
\begin{center}
\includegraphics[angle=270,width=\textwidth]{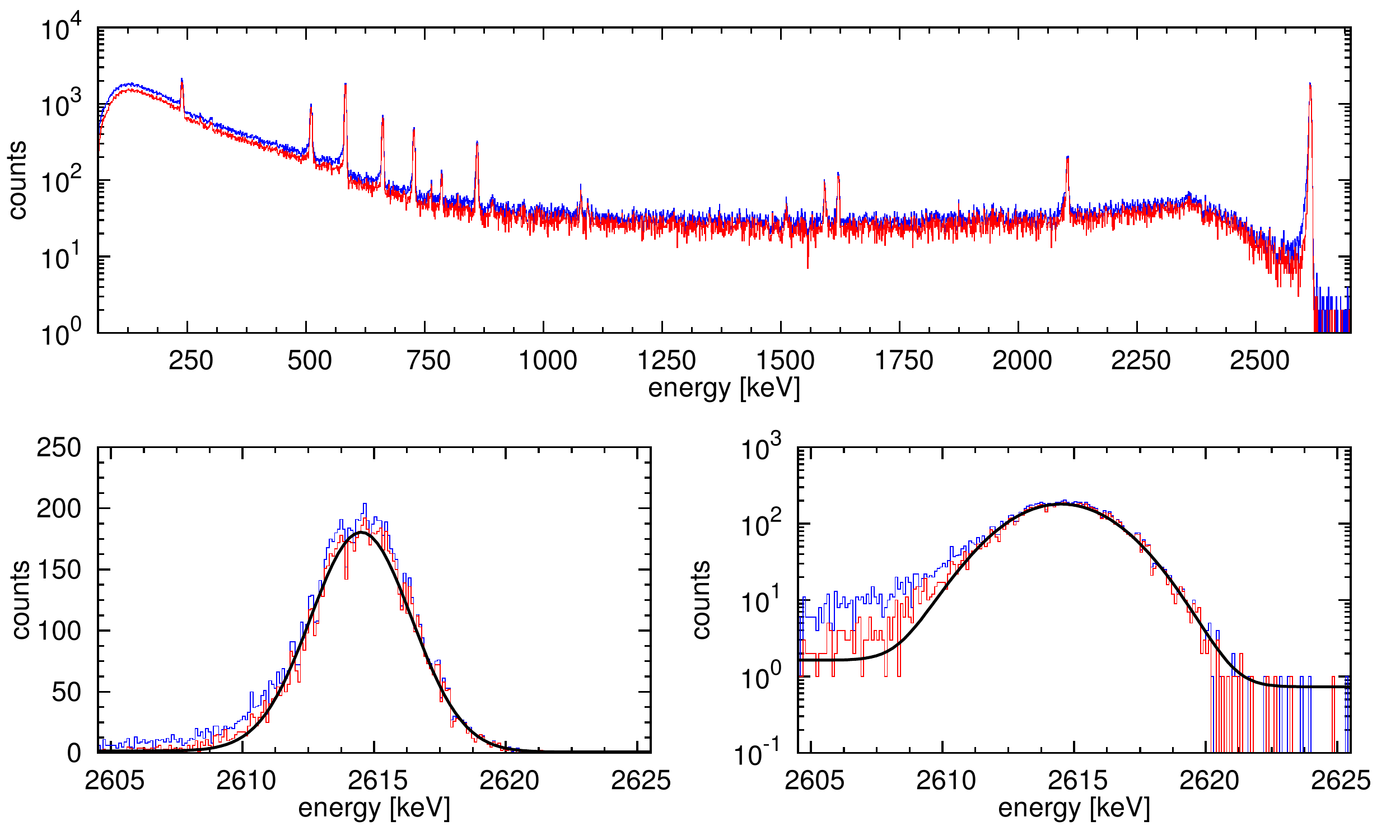}
\end{center}
\caption{\small
Reconstructed energy spectrum of an enriched detector deployed in the \gerda\
set-up irradiated with a $^{228}$Th source. The spectrum is shown before (blue
histogram) and after (red histogram) having applied the cuts described in
section~\ref{filter}. The cuts remove approximately 15\% of the total events
($\sim$10\% in the $\gamma$ lines). The bottom panels show the peak of the
2.614\,MeV $\gamma$-ray line and the analytical model used for the fit, in
linear and logarithmic scale. The tail on the left-hand side is due to pile-up
events and is visibly reduced by the cuts.
}
\label{spectrum}
\end{figure}

The results of the data selection performed according to the criteria 
described above are shown in \figurename~\ref{spectrum}.
The figure shows the energy spectrum of a HPGe detector operated in
the \gerda\ set-up irradiated with a $^{228}$Th source.
The cuts remove efficiently bad signals and pile-up events, improving the
shape of the $\gamma$-ray peaks and the agreement with the standard
analytical functions used to model $\gamma$-line peaks.

\section{Conclusions}\label{last}
The \gerda\ experiment is currently starting the data taking of Phase\,I.
The off-line analysis of the HPGe detector signals will be performed with the
\gelatio\ software framework, a tool specifically developed for \gerda, which
supports a multi-channel analysis and implements a signal processing based on a
modular approach. 
A reference analysis pipeline has been defined and optimized.
The signal
processing is performed along chains of modules and includes the estimate
of the trigger position, of the amplitude and of several 
basic pulse shape analysis parameters.
The digital filters have been improved and optimized during the \gerda\
commissioning phase and the shaping parameters have been tuned.
Also, a set of cuts has been defined to identify signals
induced by non-physical events or signals which are not properly processed.
In addition, a set of parameters was identified to monitor
the data quality and possibly to discriminate corrupted data.

The software and the digital filters have been validated during the \gerda\
commissioning and in several R\&D activities related to the experiment.
The new pipeline has been tested and it proved to be stable and ready to be used
for the reference analysis of \gerda\ Phase\,I data.

\ack
We would like to acknowledge our colleagues of the \gerda\ Collaboration,
especially D.~Budj\'{a}\v{s} and B.~Schwingenheuer, for many invaluable advices
concerning signal analysis and data quality control.
We want also to thank C. A. Ur, D. Bazzacco  and T. Kihm for all the stimulating
discussions and suggestions concerning digital signal processing and the
algorithms used in $\gamma$-ray spectroscopy.

This work was supported in part by the Transregio Sonderforschungsbereich
SFB/TR27 ``Neutrinos and Beyond'' by the Deutsche Forschungsgemeinschaft and by
the Munich Cluster of Excellence ``Origin and Structure of the Universe''.


\section*{References}
\begin{thebibliography}{99}
\bibitem{gerda} 
GERDA Collaboration, Abt I \emph{et al.}
2004
GERDA: The GERmanium Detector Array for the Search of Neutrinoless $\beta
\beta$ Decay of $^{76}$Ge at LNGS
\emph{Laboratori Nazionali del Gran Sasso - Proposal}
%
\nonum 
GERDA Collaboration, Sch\"onert S \emph{et al.}
2005
The GERmanium Detector Array (GERDA) for the search of neutrinoless beta beta decays of Ge-76 at LNGS
\emph{Nucl. Phys. Proc. Suppl.} B
\textbf{145}
242-245
%
\bibitem{hm}
Gunther M \emph{et al.}
1997
Heidelberg - Moscow beta-beta experiment with Ge-76: Full setup with five
detectors
\emph{Phys. Rev.} D
\textbf{55}
54-67
%
\nonum
Klapdor-Kleingrothaus H V, Krivosheina I V, Dietz A and Chkvorets O 
2004
Search for neutrinoless double beta decay with enriched Ge-76
in Gran Sasso 1990-2003
\emph{Phys. Lett.} B 
\textbf{586}
198-212
%
\nonum
IGEX Collaboration, Aalseth C E \emph{et al.}
2002
The IGEX Ge-76 neutrinoless double beta decay experiment: Prospects for next
generation experiments
\emph{Phys. Rev.} D
\textbf{65}
092007
%
\bibitem{bege}
CANBERRA Broad Energy Ge (BEGe) Detector, 
URL http://www.canberra.com/products/485.asp
%
\bibitem{psd} 
Budj\'a\v{s} D, Barnab\'e Heider M, Chkvorets O, Khanbekov N and Sch\"onert S
2009
Pulse shape discrimination studies with a Broad-Energy Germanium detector for
signal identification and background suppression in the GERDA double beta decay
experiment
\emph{JINST}
\textbf{4}
P10007
(preprint arXiv:0909.4044)
%
\nonum
Agostini M \emph{et al.}
2011 
Signal modeling of HPGe detectors with a small read-out electrode and
application to neutrinoless double beta decay search in Ge-76
\emph{JINST} \textbf{6} P03005 
(preprint arXiv:1012.4300)
%
\nonum
Gonzalez D \emph{et al.}
2003
Pulse shape discrimination in the IGEX experiment
\emph{Nucl. Instrum. Meth.} A
{\bf 515} 
634-643
%
\nonum
Hellmig J and Klapdor-Kleingrothaus H V  
2000
Identification of single-site events in germanium detectors by digital pulse
shape analysis
\emph{Nucl. Instrum. Meth.}  A 
{\bf 455}
638-644
%
\bibitem{gelatio}
Agostini M, Pandola L, Zavarise P and Volynets O 
2011 
GELATIO: a general framework for modular digital analysis of high-purity
Ge detector signals
\emph{JINST}
\textbf{6}
P08013
(preprint arXiv:1106.1780)
%
\bibitem{mgdo}
Agostini M \emph{et al.}
2011
The MGDO software libraries for Germanium neutrinoless double beta decay experiments
\emph{J. Phys.: Conf. Series}
 to appear in TAUP 2011
%
\bibitem{daq}
MIZZI Computer Software GmbH, URL www.mizzi-computer.de
%
\nonum
Isocrate R \emph{et al.} 
2004
MD2S - Digital sampling electronics for the MARS detector,
A data acquistion system for the MD2S digital sampling electronics
\emph{Laboratori Nazionali di Legnaro - Annual Report}
226-227

\nonum
Ur C A \emph{et al.}
2004
A data acquistion system for the MD2S digital sampling electronics
\emph{Laboratori Nazionali di Legnaro - Annual Report}
228-229
%
\bibitem{dspguide}
Smith S W 
1999 
\emph{The Scientist and Engineer's Guide to Digital Signal
Processing}
(California: California Technical Publishing San Diego)
%
\end{thebibliography}
\end{document}